# Benefits of Semantics on Web Service Composition from a Complex Network Perspective

Chantal Cherifi[1,2], Vincent Labatut[1] and Jean-François Santucci[2],

[1] Galatasaray University, Computer Science Department, Ortaköy
Istanbul, Turkey
[2] University of Corsica, UMR CNRS, SPE Laboratory, France
vlabatut@gsu.edu.tr

**Abstract.** The number of publicly available Web services (WS) is continuously growing, and in parallel, we are witnessing a rapid development in semantic-related web technologies. The intersection of the semantic web and WS allows the development of semantic WS. In this work, we adopt a complex network perspective to perform a comparative analysis of the syntactic and semantic approaches used to describe WS. From a collection of publicly available WS descriptions, we extract syntactic and semantic WS interaction networks. We take advantage of tools from the complex network field to analyze them and determine their properties. We show that WS interaction networks exhibit some of the typical characteristics observed in real-world networks, such as short average distance between nodes and community structure. By comparing syntactic and semantic networks through their properties, we show the introduction of semantics in WS descriptions should improve the composition process.

## 1 Introduction

A Web Service (WS) is an autonomous software component which can be published, discovered and invoked for remote use. For this purpose, its characteristics must be made publicly available, under the form of a so-called service description. This file comparable to interfaces defined in the context of object-oriented programming lists the operations implemented by the service. Currently, production WS use syntactic descriptions expressed with the WS description language (WSDL), which is a W3C (World Wide Web Consortium) recommendation. Such descriptions basically contain the names of the operations and their parameters names and data types, plus some lower level information regarding the network access to the service. WS were initially designed to interact with each other, in order to provide a composition of WS able to offer higher level functionalities [1]. Current production discovery mechanisms support only keyword-based search in WS registries and no form of inference nor flexible match can be performed [2]. More advanced research (non-production) approaches rely on comparing structured data such as parameters types and names, or analyzing unstructured textual comments [3-5]. This is generally not enough to distinguish WS in terms of functionality, and consequently makes it difficult, or even impossible, to use these methods to automate WS composition. Indeed, syntactically discovered WS must be manually validated to ensure they implement the desired behavior, leading to static, *a priori* compositions.

To solve this limitation, the WS research community introduced semantics in WS descriptions, through the use of new semantic description languages. Different formats exist, among which we can distinguish purely semantic descriptions (OWL-S, a W3C recommendation), from annotated WSDL descriptions (WSDL-S and SAWSDL). Although those languages allow to associate ontological concepts with various elements of the description, the research community has been focusing only on the concepts qualifying the operations inputs and outputs. Retrieving semantic information is far more costly than collecting syntactic descriptions, even when considering only parameters. The latter can be performed quickly and completely automatically. The former is a long task, requiring human intervention to label each parameter with the proper concept. Annotation tools exist to help, but they are clearly not mature yet, and often defined for specific collections or languages [6, 7]. Maybe for these reasons, no semantic annotation language emerged as an industry standard. Although they appeared more than five years ago, all production WS still rely on WSDL. Even at a research level, very few publicly available significant collections of semantically annotated WS exist, making it very difficult to test new algorithms.

This situation leads to one question: is describing WS semantically worth the cost? To our knowledge, no one did ever compare the information underlying syntactic and semantic WS descriptions. In this work, we try to tackle this problem from the service composition perspective, through the use of complex networks. We consider a set of WS as a broad interaction space, in which WS are related if they can be chained up in a composition process. We model this space by building so-called interaction networks, based on syntactic and semantic descriptions of a given WS collection. We assume that the information conveyed by the two different kinds of descriptions appears in the corresponding interaction networks. We then compare the syntactic and semantic descriptive approaches through the networks topological properties. Our main contributions are the formal definition of three types of semantic networks, an extended investigation of the WS networks topology and the comparison of syntactic and semantic networks. In section 2, we present complex networks and their main topological properties. Section 3 introduces interaction networks and explains how they can be extracted from WS descriptions. Section 4 is dedicated to the presentation and discussion of our experimental results, i.e. the obtained networks, their topological properties and how they compare. Finally, in section 5, we emphasize the original points of our work, discuss its limitations and their possible solutions.

## 2 Complex Networks Properties

Complex networks are a specific class of graphs, characterized by a huge number of nodes and non trivial topological properties. Used in many different fields to model real-world systems [8], they have been intensively studied both theoretically and practically [9]. Because of their complexity, specific tools are necessary to analyze and compare them. This is usually performed through the comparison of several well-known properties, supposed to summarize the essential of the network structure.

The *distance* between two nodes is defined as the number of links in the shortest directed path connecting them. At the level of the whole network, this allows to process the *average distance* and the diameter. The former corresponds to the mean distance over all pairs of nodes [9]. This notion is related to the *small world* property, observed when this distance is relatively small. The classic procedure to assess this property consists in comparing the average distance measured in some network of interest to the one estimated for an Erdős–Rényi (ER) network [10] containing the same numbers of nodes and links, since this random generative model is known to produce networks exhibiting the small world property [9]. In terms of dynamic processes, the existence of shortcuts between nodes can be interpreted as propagation efficiency [11]. Most real-world networks have the small world property. The *diameter* is the greatest distance over all pairs of nodes in the network.

A network *transitivity* (also called clustering) corresponds to its density of triangles, where a triangle is a structure of three completely connected nodes. It is measured by a transitivity coefficient, which is the ratio of existing triangles to possible triangles in the considered network [11]. The higher this coefficient, the more probable it is to observe a link between two nodes which are both connected to a third one. A real-world network is supposed to have a higher transitivity than the corresponding ER network by an order of magnitude corresponding to their number of nodes, meaning their nodes tend to form densely connected groups.

The *degree* of a node corresponds to the number of links attached to it. In a directed network, one can distinguish in and out degrees, i.e. the numbers of incoming and outgoing links, respectively. Nodes with a high in (resp. out) degree are called authorities (resp. hubs). The degree distribution of a network is particularly revealing of its structure. Most real-world networks have a power law degree distribution [9, 12, 13], resulting in the so-called *scale free* property. In other words, real-world networks contain a very few nodes with extremely high degree, and a massive number of nodes with very small degree.

A *component* is a maximal connected sub graph, i.e. a set of interconnected nodes, all disconnected from the rest of the network. The component distribution and, more specifically, the size of the largest component are important network properties. Indeed, depending on the context, the fact the network is split in several separated parts with various sizes can be a direct representation of the modeled system effectiveness at doing its job. For example, in a communication network like the Internet, the size of the largest component represents the largest fraction of the network within which communication is possible [9]. Most real-world networks have a so-called giant component, whose size is far greater than the other components.

A *community* is defined as a subset of nodes densely interconnected relatively to the rest of the network. Unlike components, communities are not necessarily disconnected from each other (and generally, they are significantly connected). Specific community detection algorithms must be used to identify them, leading to a partition of the overall nodes set. Most of them are dedicated to undirected networks, and only a very few recent tools can use the information conveyed by directed links. We chose to use a well tested program, and therefore focused on undirected links. We selected the Walktrap algorithm which exhibits good performances according to recent benchmarks [14]. To assess the quality of a network partition, the standard measure is Newman's modularity [15], whose value also depends on the considered network structure. Consequently, its theoretical maximal value of 1 (perfect community structure and partition) is rarely reached, and in practice values between 0.3 and 0.7 are considered high [16]. A value of 0 represents a random partition or the absence of community structure. Many real-world networks have a community structure [9].

## 3   Interaction Networks

Generally speaking, we define an interaction network as a directed graph whose nodes correspond to interacting objects and links indicate the possibility for the tail nodes to act on the head nodes. They can be considered as complex networks, and few authors used similar approaches to model collections of WS, using different granularity levels [17] and based on syntactic [17, 18] or on semantic [19-23] descriptions. In this work, we focused on networks of operations, because operations are the main point of interest when it comes to WS composition. We used both syntactic and semantic descriptions, since our goal is to compare the two types of WS descriptions.

As stated before, a WS interface is defined under the form of a set of operations. An operation $i$ represents a specific functionality, described independently from its implementation, for interoperability purposes. Besides its functionality, it is characterized by two sets of input and output parameters, noted $I_i$ and $O_i$, respectively. In a syntactic description, each parameter has a name and a type. This type is also defined independently from any implementation, again for interoperability reasons. Most of the time, the XML schema definition language (XSD) is used. In a semantic description, name and type are also generally specified, and an additional ontological concept is associated to the parameter, in order to give it a precise meaning. The most popular language used to describe these concepts is OWL, which also generally uses an XML representation.

To represent a collection of WS description under the form of an interaction network of operations, we first define a node to represent each operation in the collection. Then, a link is drawn from an operation $i$ towards another operation $j$ iff for each input parameter in $I_j$, a similar output parameter exists in $O_i$. In other words, the link exists if and only if operation $i$ can provide all the information requested to apply operation $j$. In Fig. 1, the left side represents a set of considered operations (numbers) and their input and output parameters (letters), whereas the right side corresponds to the associated interaction network of operations. All the second operation inputs ($I_2 = \{c, d\}$) are included in the first operation outputs ($O_1 = \{c, d, e\}$), i.e. $O_2 \subset O_1$, so a link exists between these operations in the interaction network. On the contrary, neither the first nor second ($O_2 = \{e, f\}$) operations provide all the parameters required by the third one ($I_3 = \{a, f, g\}$), which is why there is no link pointing towards the third operation in the interaction network.

In the interaction network, a link between two operations therefore represents the possibility to compose them. Determining if two parameters are similar is a complex task which depends on the nature of the considered parameters (syntactic vs. semantic description) and on how the notion of similarity is defined. These factors are implemented under the form of a so-called matching function.

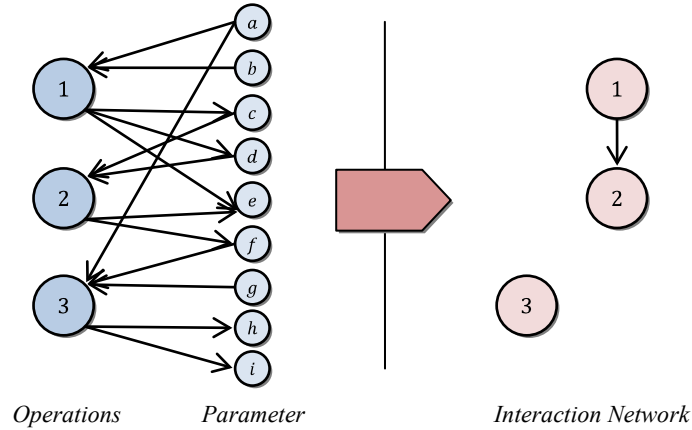

*Operations*    *Parameter*    *Interaction Network*

**Fig. 1.** Example of interaction network extraction.

A matching function $f$ takes two parameters $p_1$ and $p_2$, and determines their level of similarity [24], generally under the form of a value in $[0,1]$. It can be either symmetrical ($f(p_1, p_2) = f(p_2, p_1)$) or asymmetrical, and its output can be either binary or real. When comparing two parameters, a real output allows representing various levels of similarity, which is a good thing because it conveys a more subtle information than a raw binary value. But it also results in a more complex processing during network generation, and possibly in a network containing weighted links. Yet, most of the standard tools we decided to use to compare networks are defined for unweighted networks only, so we selected only binary matching functions in order to avoid this situation. Because of the different nature of the concerned information, we used different matching functions to compare syntactically and semantically described parameters, resulting in syntactic and semantic interaction networks, respectively. For syntactic descriptions, we compare parameters names: two parameters are said to be similar if their names are the exact same strings. The semantic matching is performed using the ontological concept associated to the parameters. We selected the three operators classically used in previous WS-related works to compare ontological concepts [25]: exact (symmetrical), plug-in and subsume (both asymmetrical). The first corresponds to a perfect matching, i.e. both concepts belong to the same ontology and are exactly identical. The second means the concept associated to the first parameter is more specific than the other one; the third represents the fact the first concept is more general than the second one. Note more flexible matching functions can be defined, both for syntactic [26] and semantic [27-30] descriptions. Our main goal is to compare syntactic and semantic descriptions, not matching functions. We opted for standard and simple tools. In summary, we can extract four distinct networks: syntactic equal $N^{Eq}$, and semantic exact $N^{Ex}$, plug-in $N^{Pg}$ and subsume $N^{Sb}$.

## 4 Results and Discussion

We extracted interaction networks from the SAWSDL-TC1 collection of WS descriptions [31, 32]. This test collection provides 894 semantic WS descriptions written in SAWSDL, and distributed over 7 thematic domains (education, medical care, food, travel, communication, economy and weapon). It originates in the OWLS-TC2.2 collection, which contains real-world services descriptions retrieved from public IBM UDDI registries, and semi-automatically transformed from WSDL to OWL-S. This collection was subsequently resampled to increase its size, and converted to SAWSDL. An SAWSDL file describes a service both syntactically and semantically. This allowed us to extract our syntactic and semantic networks from the same collection, and to compare them consistently. Other publicly available collections exist, but they do not suit our needs for different reasons. The collection must be composed of a large number of real WS, described both syntactically and semantically. The ASSAM *fulldataset* and OPOSSum *SWS-TC* [33, 34] collections are only syntactic for the first and semantic for the second. The ICEBE05 test set (not available on line anymore) is a huge collection, but its descriptions have been artificially generated. *SWS-TC* and *Jena Geography Dataset* collections from OPOSSum [34, 35] are too small to be studied through our complex networks approach.

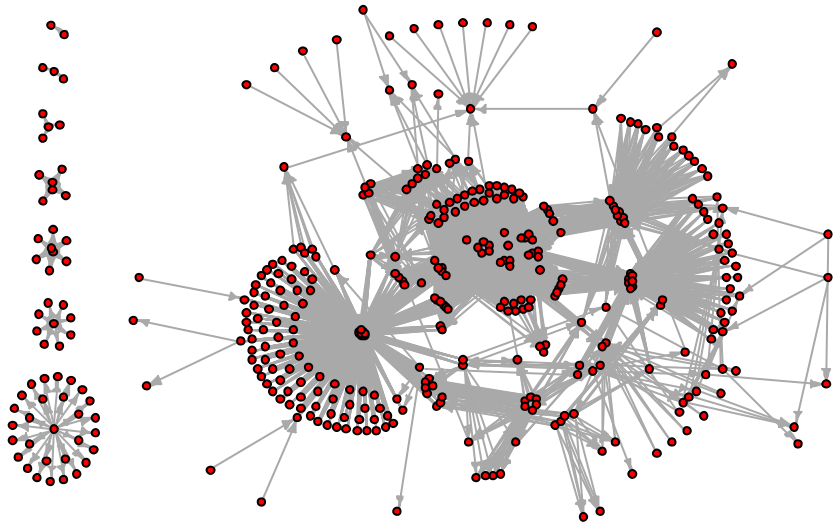

**Fig. 2.** Trimmed exact semantic network $N^{Ex}$. The giant component is located on the right side.

All the extracted networks contain many isolated nodes. They represent 44% of the total nodes in the syntactic network ($N^{Eq}$), whereas this proportion is approximately 49% in the semantic networks ($N^{Ex}$, $N^{Pg}$ and $N^{Sb}$). All networks exhibit a giant component. In the syntactic network, $N^{Eq}$, it contains 90% of the remaining nodes and 99% of the remaining links. The others 5 components are much smaller ranging from 2 to 22 nodes. The semantic network $N^{Ex}$ is separated in 8 distinct components, the giant one containing 85% of the nodes and 98% of the links in the trimmed network (see Fig. 2). The other semantic networks ($N^{Pg}$ and $N^{Sb}$) present very similar proportions. The fact distinct components exist reflects the decomposition of the collection into several non-interacting groups of operations. The presence of a giant component is a good property, because it means the number of possible interactions is high, allowing a large proportion of operations to participate in a composition. In the rest of this section, we focus on the giant components properties, discarding isolated nodes and smaller components. As shown in Table 1, the number of nodes and links is globally higher in the syntactic network, $N^{Eq}$, than in the semantic networks. This is mainly due to the fact some operation parameters have exactly the same name and are therefore considered as equal in terms of syntactic matching, whereas they do not have the same meaning. For instance, two parameters can be both named *parameter* and represent completely different data. In terms of interaction, this kind of false positive matching results in inappropriate compositions. Using semantic information is supposed to decrease this ambiguity and improve the composition process.

**Table 1.** Properties of the giant components.

| Property | Syntactic | $N^{Eq}$ | Semantic | $N^{Ex}$ | $N^{Pg}$ | $N^{Sb}$ |
|---|---|---|---|---|---|---|
| Nodes | | 395 | | 341 | 369 | 329 |
| Links | | 3666 | | 3426 | 2446 | 3864 |
| Average distance | | 2.19 | | 1.87 | 1.32 | 1.39 |
| Diameter | | 8 | | 4 | 3 | 3 |
| Transitivity | | 0.032 | | 0.022 | 0.002 | 0.027 |
| Communities | | 16 | | 12 | 7 | 4 |
| Modularity | | 0.52 | | 0.50 | 0.13 | 0.07 |

As shown in Table 1, both syntactic and semantic networks exhibit small average distances: 2.2 and 1.3~1.9, respectively. By comparison, this distance is approximately ranging from 2 to 3 in ER random networks of comparable size, which means the interaction networks possess the small-world property. In other words, many short cuts exist in the networks, indicating one can find compositions implementing a requested functionality using a relatively small number of operations. The component diameter is a good indicator of the largest possible composition, which leads here to surprisingly small values (3~8) regarding the network size. This may be due to the fact the considered collection was initially designed to assess matchmaking functions, and not really composition processes. One may also observe the significant difference between syntactic (8) and semantic networks (3~4), which confirms our previous remark regarding how their connectivity differs. More than the syntactic/semantic opposition, this seem to be related to the matching function flexibility, since both average distance and diameter are smaller for plug-in ($N^{Pg}$) and subsume ($N^{Sb}$) than for exact network ($N^{Ex}$).

Unlike most real-world networks, the measured transitivity is relatively low for both syntactic and semantic networks. Indeed, as shown in Fig.2, operations are organized hierarchically, resulting in a network structure dominated by trees rather than triangles. This favors the apparition of hubs and authorities, the former corresponding to operations possibly usable by many other operations and the latter to operations possibly using many other operations. They play a central role in the composition process, and their failure can be critical. If some operation is a hub, its output is needed by many other operations. If it becomes unavailable, all these operations cannot be composed anymore, unless other operations providing them equivalent parameters exist. If some operation is an authority, it can be composed from many other operations, which makes it less sensitive to their failure.

We fitted the networks degree distributions to the power law using the method proposed in [36]. We obtained almost zero p-values and therefore rejected this hypothesis, for all 4 networks. Even when analyzing these distributions empirically, it was difficult to interpret them, because no pattern seemed to emerge. We propose two explanations for this. First, power law fitting tests requires much data, and the size of the considered collection might be too small to allow a consistent analysis. Second, an important bias might have been introduced when the collection was resampled to increase its size, since this was certainly performed without any regard concerning the preservation of the degree distribution. However this is difficult to assess since we do not have access to this information, but only to the resulting collection. Of course, it is also possible our networks simply do not possess the scale free property.

The Walktrap algorithm detected communities with a good modularity for equal and exact networks. This community structure seems to reflect the collection domains, i.e. there is a partial correspondence between the groups of operations retrieved from the network structure and those defined thematically. Indeed, it makes sense to observe denser relationships between operations belonging to the same application field, because it is likely they were designed to be complementary and interact with each other. The low modularity observed for plug-in and subsume networks ($N^{Pg}$, $N^{Sb}$) is certainly due to the looser matching function, leading to more links between operations from different domains.

## 5 Conclusion

In this paper, we compared the information conveyed by syntactic and semantic WS descriptions, through the use of complex networks. For this purpose, we extracted 4 different interaction networks (1 syntactic and 3 semantic) from one collection of descriptions, using different matching functions. We processed, discussed and compared their topological properties. All four networks exhibit some properties observed in most real-world complex networks: small average distance, presence of a giant component and community structure. This globally reflects a large number of potential compositions. Some other properties expected from a real-world network are missing: transitivity is low and degrees are not power law distributed. We suppose this is due to the properties of the collection we analyzed, more precisely its hierarchical nature and the fact it was partially constituted by resampling.

When comparing syntactic and semantic networks, we observed the syntactic giant component was slightly larger, which might be due to the presence of false positives, i.e. operations irrelevantly connected. Although semantic giant components contain less links, their interconnection structure is more efficient, leading to a smaller average distance between operations (in terms of composition) and a smaller diameter (maximal composition size). We can conclude the introduction of semantics in WS description allows a more accurate representation of their potential interactions, and should consequently result in a more efficient search for composition processes, at least for the considered collection. When comparing only the three semantic networks, a clear distinction appears between the loose matching functions and the exact one. They lead to networks with even smaller diameters and average distances, corresponding to a larger proportion of links between the domains, which in turns result in a weaker community structure. This highlights the importance of the selected matching function.

The approach of representing a WS collection with an interaction network is generally used in the context of composition mining, i.e. to find the best composition relatively to some criteria of interest [17-23]. Oh studied some of their topological properties, but only for syntactic networks, and did not consider the directed nature of the interaction networks nor their communities [17], which are of utmost importance in the context of WS composition. Additionally, this is the first time, to our knowledge, an analysis is conducted on the topology of semantic networks, and consequently on the comparison with syntactic networks.

We can see two main limitations to our study, which we hope to solve in the near future. First, the collection we used is based on a set of real-world WS descriptions, but half of them were generated through resampling, so it cannot be considered as perfectly realistic. As a matter of fact, no other publicly available collection provides both syntactic and semantic descriptions for the same services, which is an indispensable prerequisite to a consistent comparison. The only solution we can see is to constitute our own collection, by semantically annotating a set of real syntactic descriptions. Second, we used a selected set of matching functions to extract the interaction networks. Many other functions exist, in particular more flexible syntactic distances [37] can be used to perform less strict comparisons of the parameters names. For semantic matching, more subtle subsume- and plug-in-based functions can also be derived, for instance by considering the geodesic distance between two concepts located on the same branch of an ontology. This could have significant implications on the resulting network properties, since it is directly related to the amount of false positives (nodes irrelevantly connected) and false negatives (nodes irrelevantly disconnected). Besides these improvements on data and matching functions, we plan to extend our work in two ways. First, we want to analyze in greater details the partial overlapping observed between communities and domains. It may correspond to operations shared between domains, which could be of great interest for a provider as they may be highly demanded. A related point is to test whether properties observed for the whole network are also valid for domains or sets of domains. Second, to confirm the observations we made on the network of operations, it is possible to extract and study equivalent networks at two other granularity levels [38]. We already performed the analysis of dependency networks at the parameters level [39], and plan to focus on interaction networks of whole WS very soon.